\documentclass[%
 reprint,
 amsmath,amssymb,
 aps,
]{revtex4-2}

\usepackage{graphicx}
\usepackage{dcolumn}
\usepackage{bm}
\usepackage[dvipsnames]{xcolor}

\begin{document}

\preprint{APS/123-QED}

\title{Unveiling a universal relationship between the $f(R)$ parameter and \\neutron star properties}

\author{K. Nobleson}
\affiliation{%
 Birla Institute of Technology and Science, Pilani Hyderabad Campus, Telangana, India\\
}%


\author{Tuhin Mallik}
\affiliation{
 CFisUC, Department of Physics, University of Coimbra, 3004-516 Coimbra, Portugal\\
}%
\author{Sarmistha Banik}
\affiliation{%
Birla Institute of Technology and Science, Pilani Hyderabad Campus, Telangana, India\\
}%


\date{\today}

\begin{abstract}
In recent years, modified gravity theories have gained significant attention as potential replacements for the general theory of relativity. Neutron stars, which are dense compact objects, provide ideal astrophysical laboratories for testing these theories. However, understanding the properties of neutron stars within the framework of modified gravity theories requires careful consideration of the presently known uncertainty of equations of state (EoS) that describe the behavior of matter at extreme densities.

In this study, we investigate three realistic EoS generated using a relativistic mean field framework, which covers the currently known uncertainties in the stiffness of neutron star matter. We then employ a Bayesian approach to statistically analyze the posterior distribution of the free parameter $\alpha$ of the $f(R)$ gravity model, specifically $f(R) = R + \alpha R^2$. By using this approach, we are able to account for our limited understanding of the interiors of neutron stars as well as the uncertainties associated with the modified gravity theory.

We impose observational constraints on our analysis, including the maximum mass, and the radius of a neutron star with a mass of $1.4 M_{\odot}$ and $2.08 M_{\odot}$, which are obtained from X-ray NICER observations. By considering these constraints, we are able to robustly investigate the relationship between the $f(R)$ gravity model parameter $\alpha$ and the maximum mass of neutron stars.

Our results reveal a universality relationship between the $f(R)$ gravity model parameter $\alpha$ and the maximum mass of neutron stars. This relationship provides insights into the behavior of neutron stars in modified gravity theories and helps us understand the degeneracies arising from our current limited knowledge of the interiors of neutron stars and the free parameter $\alpha$ of the modified gravity theory. 

\end{abstract}

\maketitle


\section{\label{sec:Intro} Introduction}
A supernova is triggered by the relentless pull of gravity as a massive star exhausts its nuclear fuel. The star's core implodes, undergoing a dramatic collapse, and compressing matter into an incredibly dense neutron star (NS). An NS is a compact stellar object composed primarily of neutrons. With a radius of just a few kilometers, yet a mass similar to that of the Sun \cite{baym1975neutron,heiselberg2000recent}, NS are remarkably compact objects. The cores of NS believed to contain extremely rare phases of matter \cite{Dexheimer2018}. When it comes to high density matter, many different phases or compositions may occur, including hyperons, quarks, superconducting matter, or colored superconducting matter.  Understanding their internal structure requires a deep understanding of both the behavior of matter at extreme densities and the principles of gravity. 

Extensive research is underway in the field of astrophysics to investigate the EoS of NS, which plays a crucial role in determining their fundamental properties such as mass, radius, and thermal evolution. Despite significant efforts, our current understanding of fundamental physics remains inadequate at high densities, leading to the absence of a unique EoS for NS \cite{Somasundaram2022ztm,Altiparmak2022,Gorda:2022jvk}. Challenges in obtaining precise nuclear physics experimental  data, uncertainties in the characteristics of nuclear matter, and limitations in observational data pose significant obstacles in accurately determining the EoS of NS. Nevertheless, recent advancements in multimessenger observations are providing fresh perspectives and valuable insights into the elusive EoS of these celestial objects. Several groups try to infer the EoS of NS by using astrophysical data \cite{Imam:2021dbe,Malik:2022ilb,Coughlin2019,Wesolowski:2015fqa,Landry:2020vaw,Huang:2023grj,Patra:2022yqc}. 

Conversely, one area of research that has gained significant attention in recent years is modified gravity theories. These theories propose modifications to Einstein's General Relativity (GR) to explain certain phenomena, such as the accelerated expansion of the universe or compatibility with quantum mechanics \cite{SAHNI2000,Padmanabhan2003,Copeland2007,Sotiriou2010,birrell_davies_1982}. One of the key predictions of modified gravity theories is the existence of scalar fields \cite{Babichev2010,Sotiriou2010,Clifton2012}. These scalar fields can affect the properties of a NS, such as its mass-radius relation and its moment of inertia \cite{Cooney2010,Orellana2013,capozziello2016mass}. The study of NS in the context of modified gravity theories is an active area of research, as it offers the possibility of testing the predictions of these theories against observational data \cite{Arapoglu2011,Astashenok2014,Yazadjiev2014,Nobleson2021}. In particular, some modified gravity theories predict that NS can have a larger radius than predicted by GR \cite{Yazadjiev2014,capozziello2016mass,AparicioResco:2016xcm,Llanes-Estrada:2016oom, Astashenok2019,Feng2018,Numajiri2023}.

On one hand, our limited knowledge of the constituents of the NS contributes to the degeneracy in mass-radius estimates. On the other hand, the free parameter of the modified gravity also contributes to the degeneracy in mass-radius estimates. Understanding the NS properties of different EoS within the framework of a modified gravity helps us constrain the degeneracies caused by EoS and the free parameter of the modified gravity \cite{Numajiri2022}. Observations of NS can be used to test the predictions of modified gravity theories. For example, measurements of the mass and radius of a NS \cite{Miller_2019,Raaijmakers2019,Miller2021}, tidal deformability in a coalescing binary NS merger \cite{abbott2018gw170817,LIGO2018} can be used to constrain the properties of the scalar field and to test the theory's predictions. Additionally, measurements of the moment of inertia and tidal deformability of NS can also be used to put the theory's claims about the NS compactness to the test \cite{yazadjiev2018,Raithel2019}.

Bayesian analysis \cite{gelman2013bayesian} is a very strong statistical approach that uses probability theory to make predictions and draw conclusions about the NS properties using mass, radius, and tidal deformability data. This way, we can quantify the uncertainty associated with their measurements and predictions, and improve our theoretical understanding. Bayesian analysis is routinely used in other major astrophysics problems, i.e., to analyze gravitational-wave signals \cite{Ashton2019}, properties of short gamma-ray bursts \cite{Biscoveanu2020}, test GR \cite{Keitel2019,Ashton2020,Payne2019} and to study a wide range of properties of NS \cite{Coughlin2019,HernandezVivanco2019,Biscoveanu2019}, including their masses, radii, and EoS \cite{Malik2022,Traversi2020}.

Several studies in the literature have investigated the effects of using different EoS in $f(R)$ gravity, but a comprehensive statistical analysis is yet to be done. In this study, we use the Bayesian inference method to generate a complete snapshot of the $f(R)$ model for various EoS. Our primary goal is to understand the relationships between the properties of NS and the free parameter of the $f(R)$ model while taking the currently known uncertainties of EoS into account. Our study will provide a comprehensive analysis of the relationship between $f(R)$ parameter and NS properties, which could help us understand the physics that governs their behavior. This insight can then be utilised to further constrain the parameter of modified gravity and to better understand the physics of NS. The results of our study will also be relevant in future studies of NS and other compact objects. The behaviour of NS and other compact objects may then be predicted more precisely using the knowledge gained from this.

The paper is organised as follows. In section II, a brief overview of the Tolman-Oppenheimer-Volkoff (TOV) equations in  $f(R)$ gravity in its non-perturbative form is given. We also describe the Bayesian framework used in this study.  In section III, we present an overview of the EoSs used in this work.  In section IV, we show the results obtained by numerically solving the modified TOV equations for various EoSs with different values of the free parameter $\alpha$. Finally, in the discussion section, we comment on the results of this study.

\section{Formalism}
\subsection{TOV in $f(R)$}
To derive the TOV equations, let us consider the following action (in the units of $G=c=1$):
\begin{eqnarray}\label{action}
S=\frac{1}{16\pi} \int d^{4}x \sqrt{-g} f(R) + S_{matter} 
\end{eqnarray}
where $g$ is the determinant of the metric $g_{\mu \nu}$, $f(R)$ is the functional form of Ricci scalar (in this case, $f(R)=R+\alpha R^2$, where $\alpha$ is the  free parameter), and $S_{matter}$ is the action of the matter field which is assumed to be perfect fluid. For compact objects, the metric can be assumed to be spherically symmetric as described below.
\begin{eqnarray}\label{metric}
ds^2 =-e^{2\phi(r)}dt^2 + e^{2\lambda(r)}dr^2 + r^2(d\theta^2 + sin^2\theta d\vartheta^2)
\end{eqnarray}
By varying the action with respect to $g_{\mu \nu}$, we can derive the TOV equations. 
We use non-perturbative method to derive the TOV equations that describe a static, spherically symmetric mass distribution under hydrostatic equilibrium. We introduce a new field $\Phi$ such that the scalar field $\varphi = \frac{\sqrt{3}}{2}ln\Phi$. We define $A^2(\varphi) = \Phi^{-1}(\varphi) = exp(-2\varphi/\sqrt{3})$ and $\beta(\varphi) = \frac{d ln A(\varphi)}{d \varphi} = -\frac{1}{\sqrt{3}}$. The full derivation of this formalism can be found in \cite{Yazadjiev2014,Nobleson2021,Nobleson2022}. The modified TOV equations in the non-perturbative method are as follows:
\begin{align}\label{tov1b}
\frac{d \lambda}{dr} = e^{2\lambda}\Biggl[4\pi \rho r A^4 + \frac{r e^{-2\lambda}}{2}\left(\frac{d \varphi}{dr}\right)^2 
\nonumber \\ 
+\frac{r(1-A^2)^2}{16 \alpha} - \frac{(1-e^{-2\lambda})}{2r} \Biggr]
\end{align}

\begin{align}\label{tov2b}
\frac{d \phi}{dr} = e^{2\lambda}\left[4\pi p r A^4 + \frac{r e^{-2\lambda}}{2}\Biggl(\frac{d \varphi}{dr}\right)^2 
\nonumber \\
- \frac{r(1-A^2)^2}{16 \alpha} + \frac{(1-e^{-2\lambda})}{2r}\Biggr]
\end{align}

\begin{align}\label{tov3b}
\frac{d^2 \varphi}{dr^2} = e^{2\lambda}\left[ \frac{A^2(1-A^2)}{4\sqrt{3} \alpha}- \frac{4 \pi A^4 (\rho - 3 p)}{\sqrt{3}}\right] \nonumber \\
- \frac{d \varphi}{dr}\left(\frac{d \phi}{dr} - \frac{d \lambda}{dr} + \frac{2}{r}\right)
\end{align}

\begin{eqnarray}\label{tov4b}
\frac{dp}{dr} = -(p+\rho)\left[\frac{d\phi}{dr} - \frac{1}{\sqrt{3}}\left(\frac{d \varphi}{dr}\right)\right]
\end{eqnarray}
where $\phi$ and $\lambda$ terms are taken from eqn \ref{metric}. The usual boundary conditions, i.e., regularity of the scalar field $\varphi$ at the star's core ($\frac{d\varphi}{dr}(0)=0$) and asymptotic flatness at infinity ($ \lim\limits_{r\to\infty} \varphi(r) = 0$) should be enforced. This implies that the spacetime outside the NS is not Schwarzschild spacetime. Setting $\rho = p = 0$ yields the equations defining the spacetime metric and the scalar field outside the NS. We provide the EoS for the NS matter $p = p(\rho)$ and apply the boundary conditions in order to simultaneously solve our systems of differential equations for the interior and exterior of the NS. The dimensions of the parameter $\alpha$ is in terms of $r^2_g$, where $r_g = 1.47664$ km corresponds to one solar mass.

\subsection{Bayesian estimation}
The Bayesian approach can do a comprehensive statistical analysis of a model's parameters for a given set of data. It provides the joint posterior distributions of model parameters, allowing one to investigate the distributions of given parameters and the correlations between them. The joint posterior distribution of the parameters $P(\Theta|D)$ based on the Bayes theorem \cite{gelman2013bayesian} can be written as 
\begin{equation}
    P(\Theta|D)=\frac{\mathcal{L}(D|\Theta) P(\Theta)}{\mathcal{Z}}
	\label{eq:bayes}
\end{equation}
where $D$ and $\Theta$ are the data and set of model parameters respectively. Here $P(\Theta)$ is the prior for model parameters, $\mathcal{L}(D|\Theta)$ is the likelihood function and $\mathcal{Z}$ is the evidence. The posterior distribution was evaluated by  Pymultinest \cite{Buchner2014} implementation.  

\subsubsection{The prior}
Our chosen model of $f(R)$ has only one parameter, $\alpha$.  We have taken a uniform distribution [3.5, 2000] to determine the prior on $\alpha$. The models with $\alpha < 3.5$ and $\alpha > 2000$ are very close to the cases of $\alpha = 0$ and $\alpha = 2000$, respectively. Similar behaviour for $\alpha$ is also reported by \cite{Yazadjiev2014}.

\subsubsection{The fit data}
The constraints used to fit for the parameter of $f(R)$ model is based on the NS observational properties, such as maximum mass ($M_{max}$), radius at maximum mass ($R_{2.08}$), and at 1.4$M_{\odot}$ ($R_{1.4}$) listed in Table \ref{tab:tab_const} 

\subsubsection{The Log-Likelihood}
The log-likelihood for mass, radius at maximum mass, and radius at $M_{1.4}$ are defined as follows:
\begin{equation}
    log M_{max}=log \bigg[\frac{1}{exp\bigg[\frac{M_{cal}-M_{obs}}{\Delta M}\bigg]+1} \bigg]
	\label{eq:masslog}
\end{equation}

\begin{equation}\begin{split}
    log R_{2.08}&=-0.5 \bigg[ \frac {R_{2.08 calc}-R_{2.08 obs}}{\Delta R_{2.08}}\bigg]^2 \\  
    &+ log\big[2 \pi \Delta R_{2.08}^2\big]
	\label{eq:r2.08log}    
\end{split}
\end{equation}

\begin{equation}\begin{split}
    log R_{1.4}&=-0.5 \bigg[ \frac {R_{1.4 calc}-R_{1.4 obs}}{\Delta R_{ 1.4}}\bigg]^2 \\
    &+ log\big[2 \pi \Delta R_{ 1.4}^2\big]
	\label{eq:r1.4log}
\end{split}
\end{equation}

\begin{ruledtabular}
    \begin{table}
	\centering
	\caption{The constraints imposed in the Bayesian inference: Observed maximum mass of NS, Radius of 2.08 $M_{\odot}$ NS, Radius of 1.4 $M_{\odot}$ NS.}
	\label{tab:tab_const}
	\begin{tabular}{lccr} 
		  &  \textbf{Constraints}&  &  \\
        Quantity & Value/Band & Reference\\
		\hline
		$M_{max}$ & $>$ 2.0 $M_{\odot}$ & \cite{Miller2021}\\
		$R_{2.08}$ & $12.4 \pm 1.0$ km & \cite{Miller2021}\\
		$R_{1.4}$ & $13.02 \pm 1.24$ km & \cite{Miller_2019}\\
	\end{tabular} \\ 

 \footnotemark[1]{Note: Please note that the NS maximum mass does not affect the likelihood in our case, and it has been included for completeness only. This is because the f(R) parameter, denoted as $\alpha$, can only increase the mass of NS. Furthermore, the EoS that we have chosen already predict NS with maximum masses above 2 M$_\odot$ in GR ($\alpha=0$). }
\end{table}
\end{ruledtabular}
The GR maximum mass for DD2, SFHx, and FSU2R are  $2.40 M_{\odot}$, $2.13 M_{\odot}$, and $2.06 M_{\odot}$, respectively. $\Delta$R and $\Delta$M represent the uncertainty in the measurement of mass and radius from the observations.

\section{Equations of State}
In this study, we focus on the EoS using only nucleonic degrees of freedom to investigate the impact of $f(R)$ gravity versus GR. There are several EoS available which make use of  different approaches such as piecewise-polytropic, constant speed of sound, Taylor expansion around the nuclear matter point, relativistic mean field (RMF) model, etc. Relativistic models are always causal, which means that the speed of sound is always slower than the speed of light. These models are tractable because of the potential for adding various interaction terms. However, the couplings are fixed by the properties of nuclear matter at saturation density, such as binding energy, or the symmetry energy and its slope L, etc. One may question the validity of the couplings at higher densities, which are typical of NS interiors. One particular advantage of these models is that they can be extended to the suitable range of temperatures and proton fractions, relevant for NS merger. We have selected three different models for the nuclear EoS, all of which are based on a RMF framework: SFHx \cite{Steiner:2012rk}, FSU2R \cite{Tolos:2016hhl}, and DD2 \cite{Hempel2010,Typel:2009sy}.  These are based on the covariant field-theoretical approach to hadronic matter \cite{Serot:1984ey,Serot1997}. Table \ref{tab:nuc_prop} provides a list of the nuclear saturation properties for each of these models.

\begin{figure}
    \centering
    \includegraphics[width=0.95\columnwidth]{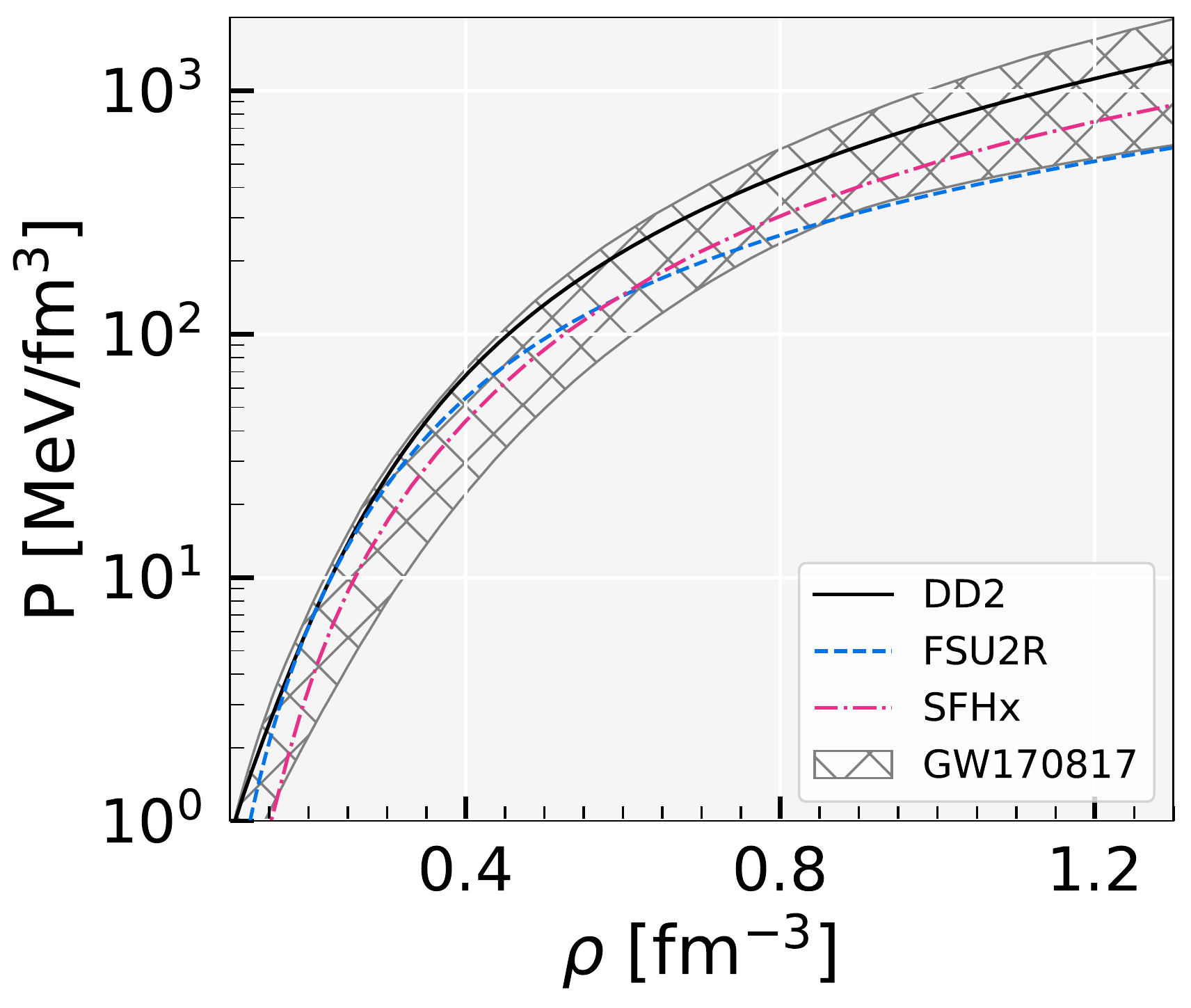}
    \caption{Pressure versus the baryonic number density for the three different RMF EoS employed for this work: DD2 (black), FSU2R (blue),  and SFHx (pink). There is also a band (hatched gray) that has been predicted based on the GW170817 event.}
    \label{fig:eos}
\end{figure}

{\it SFHx --} In the SFHx \cite{Steiner:2012rk} EoS, the lagrangian is based on the interchange of isoscalar-scalar $\sigma$, isoscalar-vector $\omega$, and isovector-vector $\rho$-mesons. Fits from the experimental data are required to estimate the free parameters in the lagrangian. It is based on an interpolation of two parameter sets, TM1 and TM2 \cite{Sugahara1994}, which were fitted to binding energies and charge radii of light (TM2) and heavy nuclei, respectively (TM1). To have a fair description of nuclei throughout the full mass number range, the coupling parameters $g_i$ of the set TMA are chosen to be mass-number dependent of the form $g_i = a_i + b_i/A^{0.4}$, with $a_i$ and $b_i$ being constants. The couplings become constants for uniform nuclear matter and are given by $a_i$.

{\it FSU2R --} The nucleonic EoS is derived as a new parameterization of the nonlinear realisation of the RMF model. Beginning with the current RMF parameter set FSU2 \cite{Chen2014}, if the pressure of NS matter in the vicinity of saturation is reduced, it allows for smaller stellar radii while maintaining nuclear matter and finite nuclei properties. Furthermore, the pressure at high densities are preserved consistent with high-energy heavy-ion collisions findings and sufficiently stiff to support 2M$_{\odot}$ NS \cite{Danielewicz2002}. 

{\it DD2 --} The basic relativistic lagrangian has effective interaction via contributions from  $\sigma,~\omega$, and $\rho$ mesons without any self-coupling factors. The density-dependent couplings allow the pressure term to rearrange and account for the system's energy-momentum conservation and thermodynamic consistency \cite{Hempel2010,Banik2002}. The DD2 model satisfies the constraints on nuclear symmetry energy and its slope parameter, as well as the incompressibility from the nuclear physics experiments \cite{Char2014}. As emphasised in \cite{Fortin2016}, proper core-crust matching is critical to avoiding uncertainty in the macroscopic properties of stars. The DD2 EoS uses the same lagrangian density to describe both the low-density crust and the high-density core, allowing for a smooth transition between the two.

In figure \ref{fig:eos}, we plot the pressure versus the baryonic number density for the three RMF EoS used in this study: DD2 (black, solid line), FSU2R (blue, dotted line), and SFHx (pink, dashed line). The grey band (hatched grey) is the prediction from the GW170817 event \cite{LIGOScientific:2018hze}. These EoS satisfy the maximum mass and radius constraints from the observation \cite{Riley:2019yda, Miller:2019cac, Riley:2021pdl, Miller:2021qha}. 

\begin{ruledtabular}
\begin{table}[htb]
	\centering
	\caption{For the EoS model employed in the work, namely SFHx \citep{Steiner:2012rk}, FSU2R \citep{Tolos:2016hhl}, and DD2 \citep{Hempel2010,Typel:2009sy}, we compile the nuclear matter saturation properties.}
	\label{tab:nuc_prop}
	\begin{tabular}{lccccccr} 
		  EoS & $n_B^0$ & B/A & $K_0$   & $Q_0$  & $J_{0}$   & $L_{0}$\\
        &($fm^{-3}$) & (MeV) &(MeV) &(MeV)   &(MeV)  &(MeV)\\
  		\hline
       SFHx& 0.160 & -16.16 &239 &-457   &28.7  &23.2\\
       FSU2R& 0.151 & -16.28 &238 &-135   &30.7  &47.0\\
       DD2& 0.149 & -16.02 &243 &169   &31.7  &55.0\\
	\end{tabular}
\end{table}
\end{ruledtabular}

\section{Results}
The posterior probability distributions of the $f(R)$ model parameter $\alpha$ are analyzed as follows. Our Bayesian approach to estimating the parameter $\alpha$ utilizes a uniform ("un-informative") prior, as described in the section 2. To incorporate the well-known uncertainty of nuclear matter EoS, we have employed three distinct nuclear matter EoS models: DD2, FSU2R, and SFHx described in the section 3. Together, these models span the majority of the presently known range of uncertainty for dense matter NS EoS. They allow us to analyse the effect of modified gravity and its dependence, if any, on these EoS. 

\begin{figure}
        \centering
	\includegraphics[width=0.85\columnwidth]{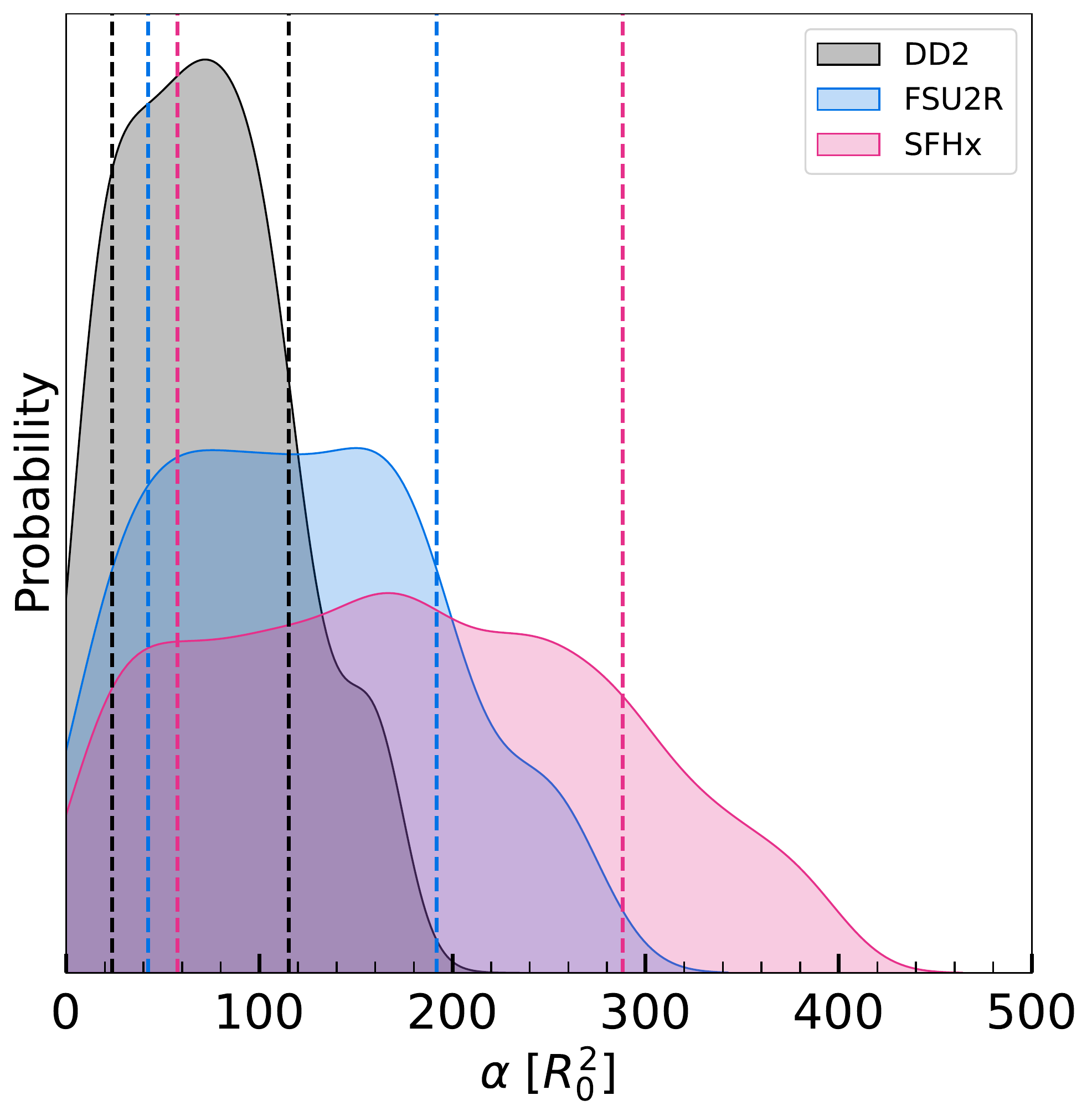}
    \caption{For three different EoS model, namely DD2, FSU2R, and SFHx, the final posteriors of the parameter $\alpha$ for $f(R)$ gravity are plotted. The vertical lines show the $1\sigma$ (68\%) credible intervals (CIs).}
    \label{fig:alfa_dist}
\end{figure}

\begin{figure}[htb]
	\includegraphics[width=0.9\columnwidth]{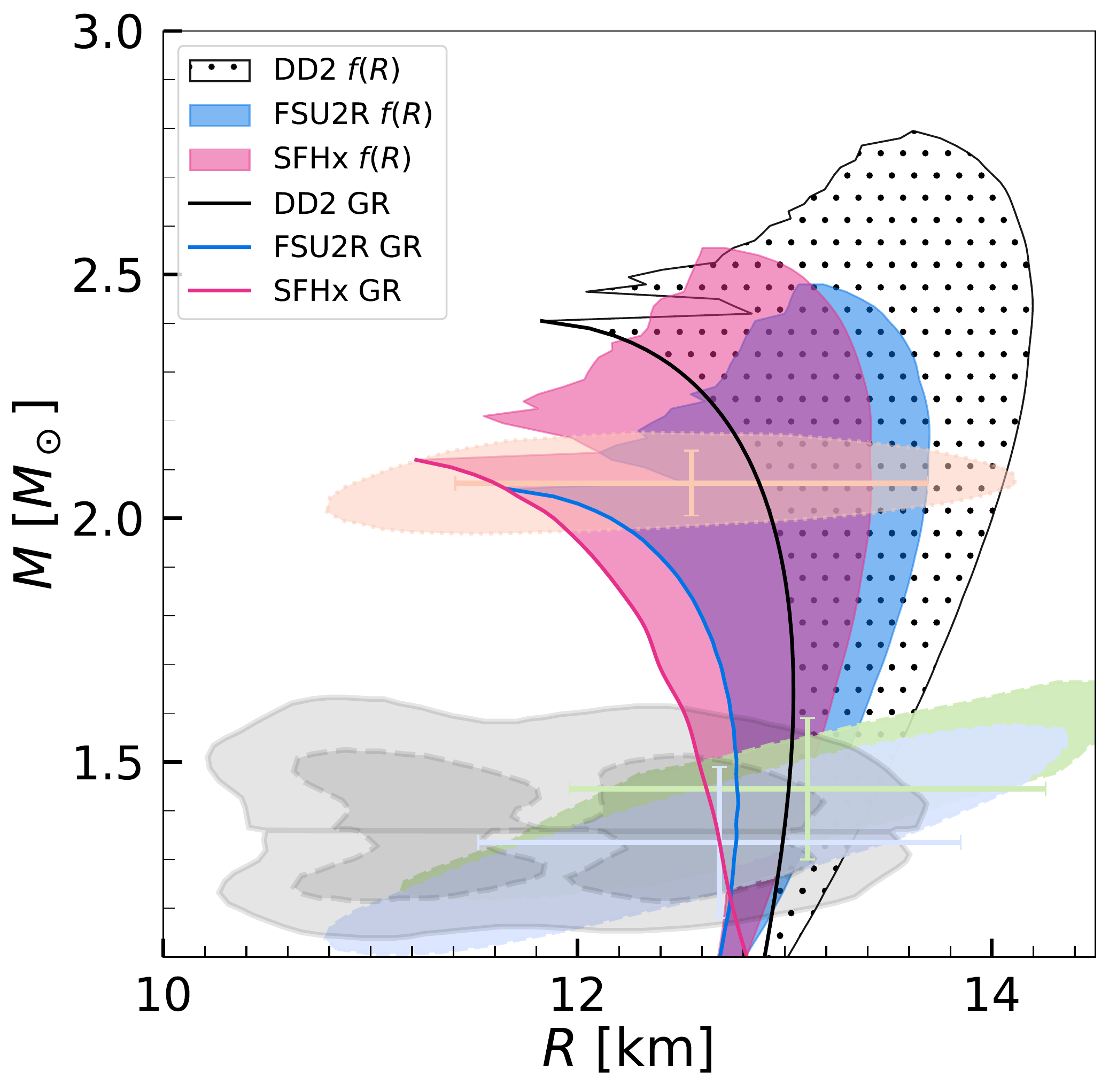}
    \caption{The entire M-R domain is plotted for three different EoS models, namely DD2, FSU2R, SFHx obtained for the full posteriors of the parameter $\alpha$ for $f(R)$ gravity. The gray zones indicate the 90\% (solid) and 50\% (dashed) credible Interval for the binary components of the GW170817 event. The $1\sigma$ (68\%) credible zone of the 2-D posterior distribution in mass-radii domain from millisecond pulsar PSR J0030+0451 (light  green and light blue) as well as PSR J0740+6620 (light orange) are shown for the NICER x-ray data. The horizontal (radius) and vertical (mass) error bars reflect the $1\sigma$ credible interval derived for the same NICER data's 1-D marginalized posterior distribution. \label{fig:alfa_MR}}
\end{figure}

In Figure \ref{fig:alfa_dist}, we present a visual representation of the posterior probability distributions for the $f(R)$ model parameter $\alpha$ obtained using three different nuclear matter EoS models. Each distribution is represented by a curve, and the vertical lines indicate the 90\% confidence interval for each case. It is interesting to observe the way the distribution's shape varies depending on the stiffness of the EoS. Specifically, the distribution is smaller (i.e., more constrained) for stiffer EoS and larger (i.e., less constrained) for softer EoS. This is due to the fact that $f(R)$ parameter $\alpha$ only increases the mass and radius of a NS. For a stiff EoS, the cost we employed for radius measurement from NICER imposes greater constraints on the value of $\alpha$, i.e., if the calculated value is very large or very small in comparison to the observed value, then such $\alpha$ value is less preferred. It is important to emphasize that the distributions of $\alpha$ are heavy-tailed, i.e., goes to zero slower than one with exponential tail. Therefore, the probability values presented in the plot are normalized to the tail of the distribution. Additionally, it is worth noting that the effect of $\alpha$ on the properties of a NS can only be observed within a certain range of values. Beyond this range, increasing the value of $\alpha$ has no impact on the star's properties (see section 2.2.1).

The full posterior of $\alpha$ was used to generate the entire mass-radius domain for three different models, namely DD2, FSU2R, and SFHx in figure \ref{fig:alfa_MR}. The grey zones in the figure's lower left corner, which represent the 90\% (solid) and 50\% (dashed) confidence intervals for the binary components of the GW170817 event \citep{LIGOScientific:2018hze}, serve as a baseline for comparison. Also, the $1\sigma$ (68\%) credible zone of the 2-D posterior distribution in the mass-radius domain obtained from millisecond pulsars PSR J0030+0451 (light green and light blue) \citep{Riley:2019yda, Miller:2019cac} and PSR J0740+6620 (light orange) \citep{Riley:2021pdl, Miller:2021qha} for the NICER X-ray data is plotted. The error bars, both horizontal (radius) and vertical (mass), represent the 1-D marginalised posterior distribution's $1\sigma$ credible interval. The figure also shows that measurement of NS radius at higher masses can greatly constrain the $f(R)$ parameter when compared to lower mass NS. It is worth noting that the $f(R)$ parameter, $\alpha$, can only broaden the MR curve in higher mass. Observations of NS with high masses can thus provide valuable insights into the fundamental nature of these objects.

\begin{figure*}
	\includegraphics[width=0.6\columnwidth]{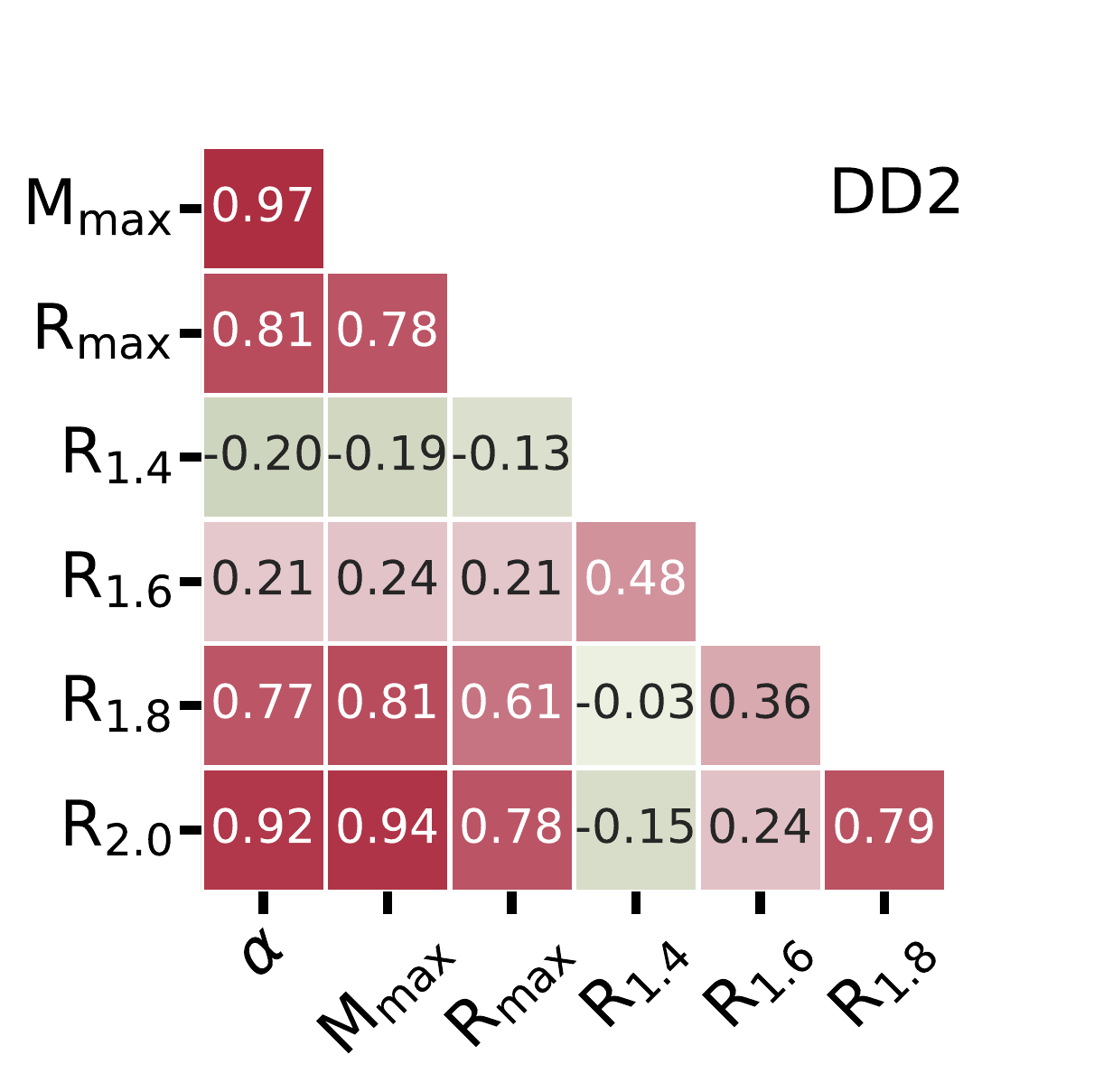}
        \includegraphics[width=0.6\columnwidth]{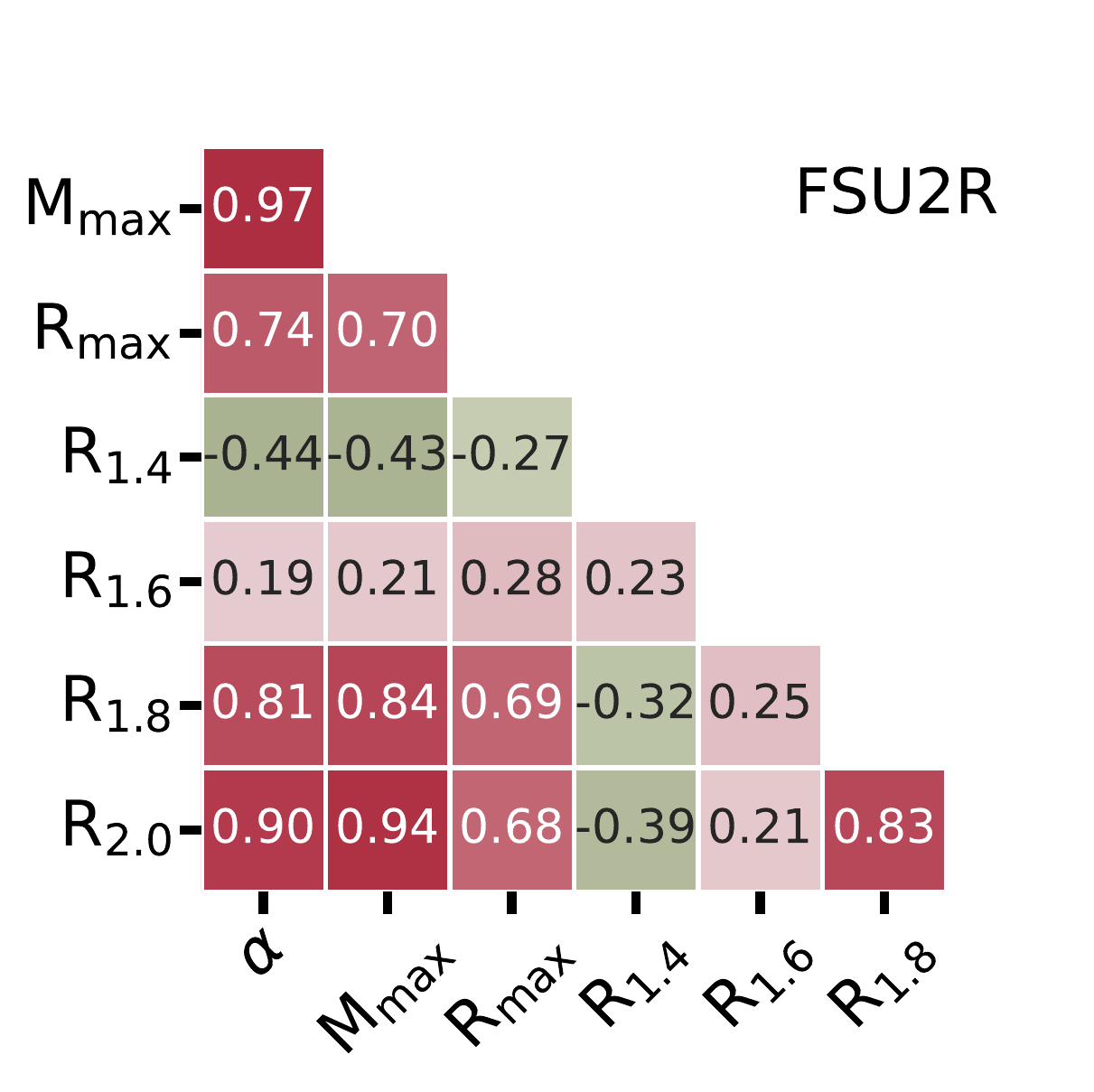}
        \includegraphics[width=0.6\columnwidth]{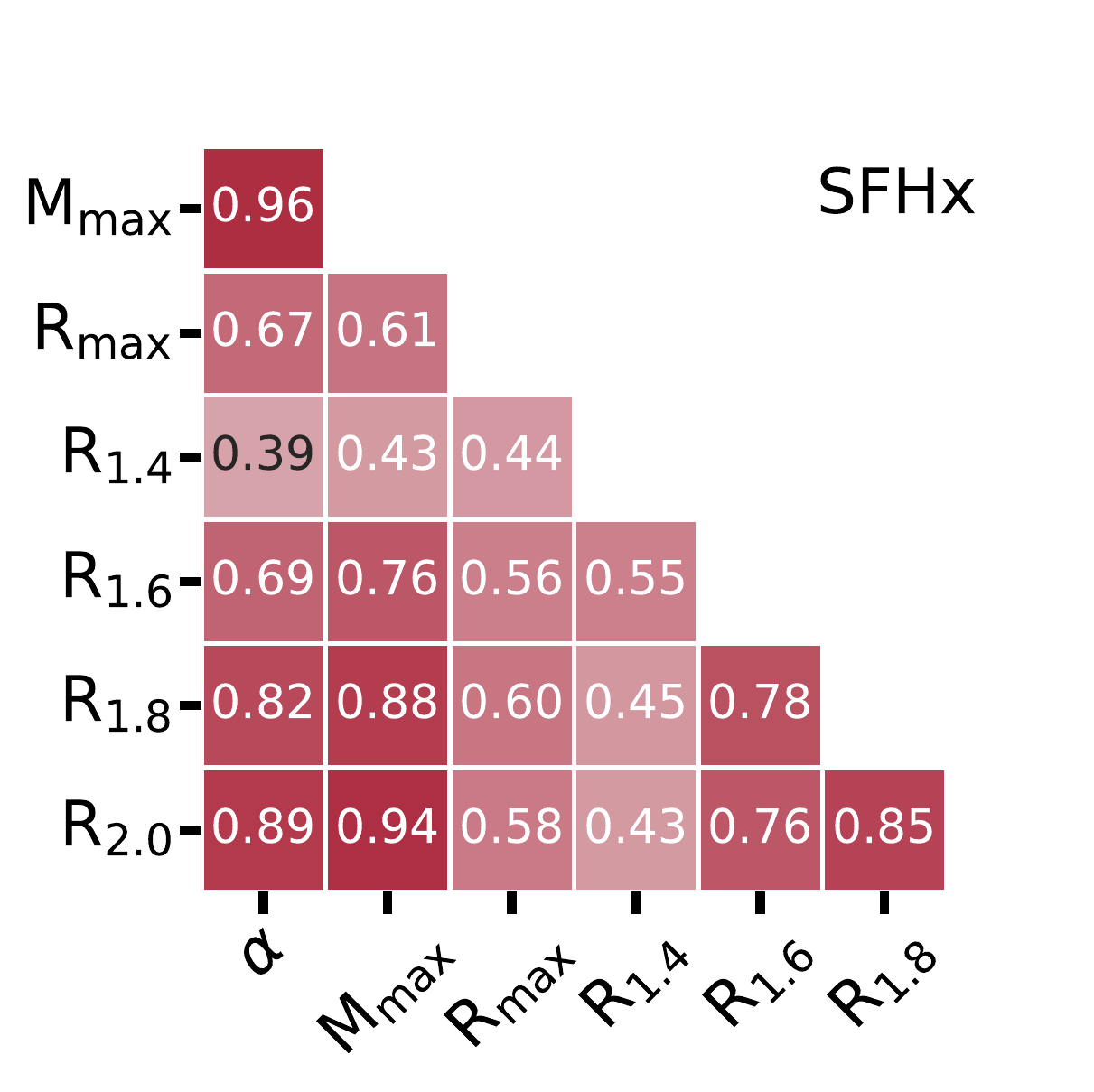}
        \caption{We plot the Kendall rank correlation coefficients between parameter $\alpha$ for $f(R)$ and NS properties of different mass ranges, such as NS maximum mass M$_{\rm max}$, maximum radius R$_{\rm max}$, and radius for 1.4, 1.6 and 1.8 M{$_\odot$} NS obtained for the final posterior of three different EOSs: (left) DD2, (middle) FSU2R, and (right) SFHx.}
        \label{fig:kendal}
\end{figure*}

In figure \ref{fig:kendal}, a Kendall rank correlation coefficient \citep{Kendall1938} is presented, which represents the correlation between the parameter $\alpha$ for $f(R)$ and NS properties for different mass ranges. This correlation coefficient is derived from the final posteriors of three different EoS: DD2, FSU2R, and SFHx, for the NS maximum mass ($M_{max}$), maximum radius ($R_{max}$), and radius for 1.4, 1.6, and 1.8 $M_{\odot}$ NS. It is noteworthy that Pearson’s correlation coefficient is typically employed in such figures to measure the linear relationship between two variables. However, Kendall’s correlation coefficient is used here as it measures a monotonic relationship between two variables. Regardless of the EoS model chosen, the results show that the parameter $\alpha$ in $f(R)$ is strongly correlated with the NS maximum mass. Furthermore, there seems to be a strong correlation between the parameter $\alpha$ and the radius of the NS as the star's mass increases. As the mass of the NS goes from 1.4 M$_{\odot}$ to 2.0 M$_{\odot}$, the Kendall rank goes from $-0.20$ to $0.92$ for DD2, $-0.44$ to $0.90$ for FSU2R, and $0.39$ to $0.89$ for SFHx. 
The use of different EoS provides a comprehensive understanding of the correlation between the $f(R)$ parameter and NS properties.

\begin{figure*}
        \centering
	\includegraphics[width=2.0\columnwidth]{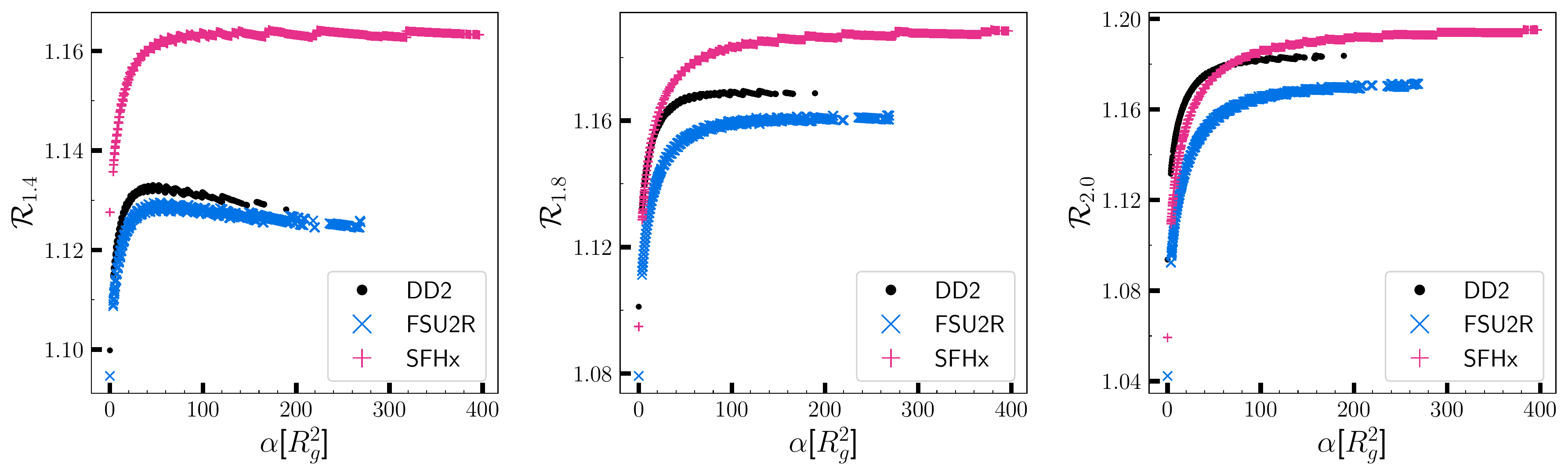}
    \caption{The dependence of normalized radius on the $f(R)$ parameter $\alpha$, as obtained from the final posterior, is shown for three different models - DD2 (black), FSU2R (blue), and SFHx (pink) - and three different NS masses, ranging from 1.4 to 2.0 M{$_\odot$} (left to right).}
    \label{fig:radius_comp}
\end{figure*}

\begin{figure*}
        \centering
	\includegraphics[width=1.95\columnwidth]{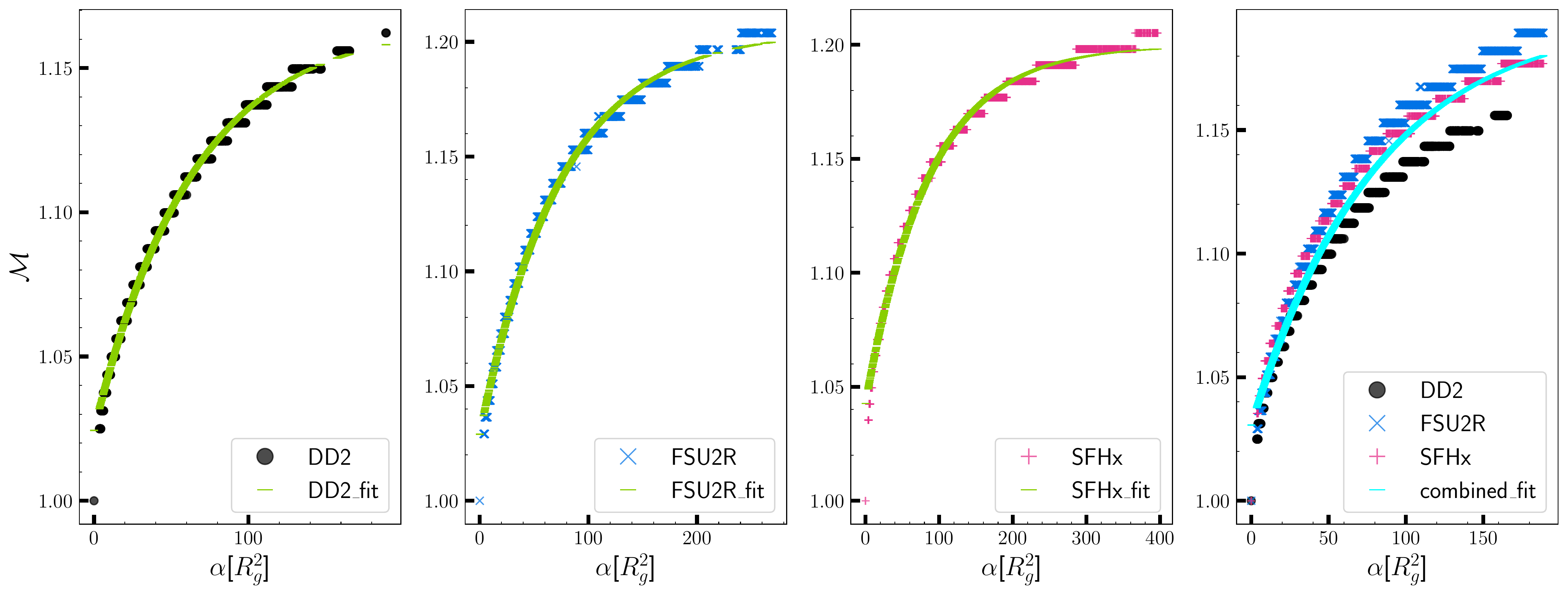}
    \caption{The dependence of normalized NS maximum mass as a function of $f(R)$ parameter $\alpha$, as obtained from the final posterior, is presented for three different models - DD2 (black), FSU2R (blue), and SFHx (pink). The green curve represents the fit curve for individual EoS. The cyan curve represents the combined fit. (see text for details).}
    \label{fig:mass_UR}
\end{figure*}

\begin{figure*}
        \centering
	\includegraphics[width=2\columnwidth]{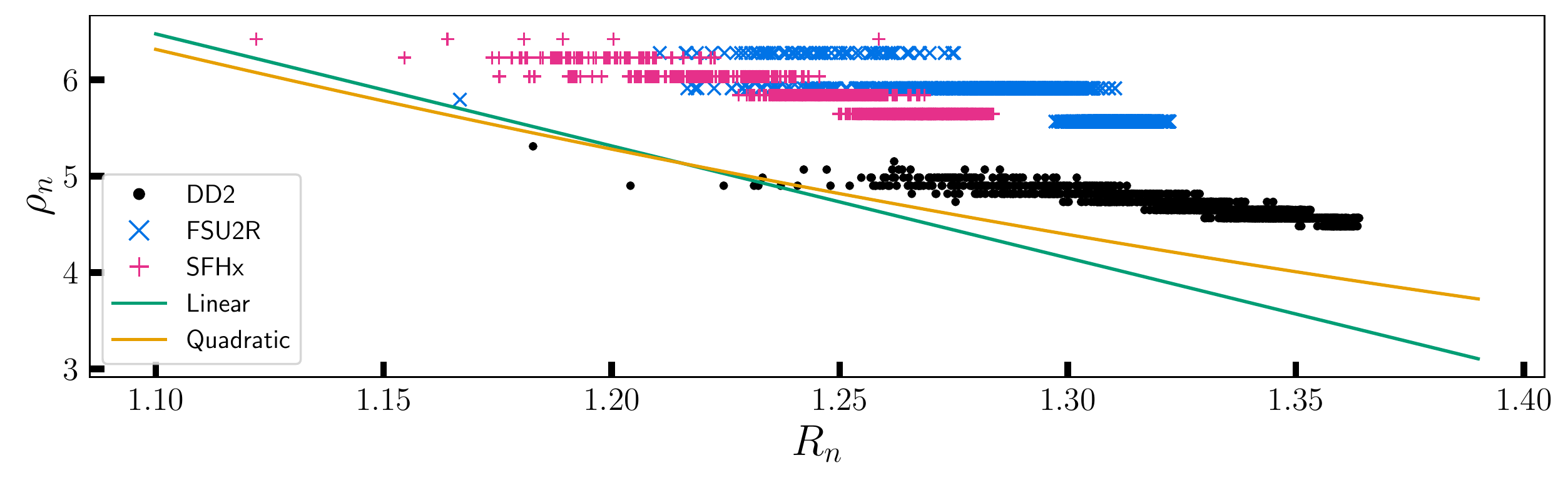}
    \caption{The correlation between normalized central density and the normalized radius of the maximum mass in $f(R)$ for the values of the free parameter $\alpha$ from the posterior for three different models - DD2 (black), FSU2R (blue), and SFHx (pink) are plotted.  The green curve represents a linear fit and orange curve represents a quadratic curve (see text for details).}
    \label{fig:rhoc_rmax}
\end{figure*}

In figure \ref{fig:radius_comp}, we define normalized radius $\mathcal{R}_n$ as $R_{n f(R)}/{R_{n GR}}$ where '$n$' is the mass of the NS, and plot $\alpha$ versus $\mathcal{R}_n$ for the three EoS models $-$ DD2 (black), FSU2R (blue), and SFHx(pink) $-$ for NS with mass of 1.4 M$_{\odot}$, 1.8 M$_{\odot}$, and 2.0 M$_{\odot}$. In the left panel of figure \ref{fig:radius_comp}, for NS with mass of 1.4 M$_{\odot}$, the normalized radius $\mathcal{R}$ increases as $\alpha$ increases up to about $\alpha = 40$. While for SFHx the value of $\mathcal{R}$ asymptotically reaches the value of 1.17, for DD2 and FSU2R the value decreases. The value of $\mathcal{R}$ is comparable for DD2 and FSU2R, whereas, for SFHx the value is significantly higher than the other EoS. In the middle panel, for NS with mass of 1.8 M$_{\odot}$, the normalized radius $\mathcal{R}$ increases as $\alpha$ increases up to about $\alpha = 50$. The value of $\mathcal{R}$ asymptotically reaches 1.16, 1.65, and 1.18 for FSU2R, DD2, and SFHx, respectively. The relative offset between the curves is reduced compared to the plot on the left. In the right panel, for NS with mass of 2.0 M$_{\odot}$, the normalized radius $\mathcal{R}$ increases as $\alpha$ increases up to about $\alpha = 50$. The value of $\mathcal{R}$ asymptotically reaches 1.17, 1.18, and 1.19 for FSU2R, DD2, and SFHx, respectively. The relative offset between the curves is minimum compared to the left and middle plots. At lower mass (left plot), there is an indication of EoS dependence of $\mathcal{R}$ versus $\alpha$. As the mass increases from 1.4 M$_{\odot}$ to 2.0 M$_{\odot}$, the EoS dependence is reduced significantly. 

In figure \ref{fig:mass_UR},
the mass and radius for each EoS are estimated with the posteriors estimated of the parameter $\alpha$.  The normalized mass $\mathcal{M}=  M_{f(R)}/{M_{GR}}$ is defined and plotted versus the parameter $\alpha$. It can be seen that any given $\mathcal{M}$ can be generated by a value of $\alpha$ and a few additional values in a small neighbourhood around it. As $\mathcal{M}$ increases, the neighbourhood around $\alpha$ also increases.
The following function is then used to fit the results: 
\begin{equation}
    \mathcal{M}(\alpha) =a + b e^{c \alpha}
	\label{eq:uni_fit}
\end{equation}
The fitting parameters for EoS DD2, FSU2R, and SFHx and the combined data are presented in Table \ref{tab:mr_fit}. $Q_{90}$ denotes the maximum percentage of relative uncertainty within a $90\%$ confidence interval. In the first subplot, The normalized mass $\mathcal{M}$ for DD2 is plotted against the parameter $\alpha$. $\mathcal{M}$ asymptotically reaches to 1.16 as $\alpha$ increases. In the second subplot, $\mathcal{M}$ versus $\alpha$ is plotted for FSU2R. $\mathcal{M}$ asymptotically reaches to 1.19 as $\alpha$ increases. In the third one, $\mathcal{M}$ versus $\alpha$ is plotted for SFHx. $\mathcal{M}$ asymptotically reaches to 1.18 as $\alpha$ increases. In the right most subplot, $\mathcal{M}$ versus $\alpha$ for all the three EoS are plotted. A curve is fitted to the combined data which is represented by the cyan curve. $\mathcal{M}$ asymptotically reaches to 1.18 as $\alpha$ increases. It is interesting to note that the stiffer EoS has relatively smaller values for '$a$' and higher values for '$b$'. Using this universal relationship, we can estimate the mass of the NS for any given value of $\alpha$ within $Q_{90}$.

\begin{ruledtabular} 
\begin{table}[htb]
	\centering
	\caption{The fit for M-R curves are listed for the EoS DD2, FSU2R, and SFHx as well as a common fit for the combined data. The maximum percentage of relative uncertainty within 90\% confidence interval is indicated in $Q_{90}$ }
	\label{tab:mr_fit}
	\begin{tabular}{lcccr} 
		  &  &\textbf{Fit values}& &\\
        EoS & a &b & c &$Q_{90}$\\
		\hline
		DD2 & 1.165	&-0.141& -0.0156&$<1$\%\\
		FSU2R & 1.205&-0.176&-0.0134&$<1$\%\\
		SFHx& 1.200&-0.157&-0.0116&$<1$\%\\
        Combined& 1.196	&-0.165&-0.0124&$<2$\%\\
	\end{tabular}
\end{table}
\end{ruledtabular}

In earlier works \citep{Jiang2022,Malik2023}, the authors showed that in GR there is a strong model independent correlation between the maximum mass star's central density $\rho_c$ and its radius $R_{max}$ for various EoS.  We want to investigate if there are any such relationships in the $f(R)$ domain. In figure \ref{fig:rhoc_rmax}, we plot normalized radius $R_n$ obtained for various $\alpha$ versus normalized central density $\rho_n$ at which the maximum mass occurs in $f(R)$ according to the following expressions.   

\begin{equation}
\rho_n = \frac{\rho_c}{0.16 ~{\rm fm}^{-3}}, \hspace{10pt} 
R_n = \frac{R_{\rm max}}{10 ~{\rm km}}
	\label{eq:norm}
\end{equation}
The authors of \cite{Malik2023} proposed a linear relation with $m_0 = -11.618 \pm 0.018$ and $c_0 = 19.255 \pm 0.019$. In the figure it is represented by a green curve.
\begin{equation}
 \frac{\rho_c}{0.16 ~{\rm fm}^{-3}}= m_0 \bigg[\frac{R_{\rm max}}{10 ~{\rm km}}\bigg]+c_0
	\label{eq:linear}
\end{equation}
The authors of \cite{Jiang2022} proposed a quadratic relation with $d_0 = 27.6$ and $d_1 = 7.55$ with a $3.7\%$ standard deviation. This is represented by a orange curve in figure \ref{fig:rhoc_rmax}.
\begin{equation}
 \frac{\rho_c}{0.16 ~{\rm fm}^{-3}}= d_0 \bigg[1-\frac{R_{\rm max}}{10 ~{\rm km}}\bigg]+d_1 \bigg[\frac{R_{\rm max}}{10 ~{\rm km}}\bigg]^2
	\label{eq:quadratic}
\end{equation}
In figure \ref{fig:rhoc_rmax}, unlike GR, the data do not support the existence of a relationship between the variables under consideration in $f(R)$. 

\section{Conclusions}
In summary, NS are important compact objects that help in the study of the EoS of matter at high densities. Understanding NS properties fully for various EoS within a modified gravity theory is critical for evaluating the degeneracies resulting by our limited knowledge of NS interiors. We use the non-perturbative TOV equations for $f(R)$ gravity in this study and a Bayesian approach to estimate the posterior distributions of model parameters. We explore the impact of $f(R)$ gravity on the parameters of the NS of three EoS using nucleonic degrees of freedom with varied stiffness and compared it to the results of GR.

We find that the measurement of NS radius at higher masses can significantly constrain the $f(R)$ parameter as compared to lower mass NS. We also find a strong correlation between the free parameter $\alpha$ and NS mass, regardless of the EoS used. Furthermore, we see that the $\mathcal{R}$ of a low-mass NS is highly sensitive to EoS, whereas the high-mass counterparts are not. Our findings reveal a universal relationship between $\alpha$ and normalised mass $\mathcal{M}$ that allows us to estimate the maximum mass of an NS for any arbitrary $\alpha$. 

In contrast to GR, the data do not support the existence of a relationship between the maximum mass star's central density and its radius in $f(R)$.

Overall, this study discusses the significance of understanding the properties of NS in modified gravity theories and provides valuable insights into the nature of these compact objects. In future, this study may be extended to include a variety of EoS classes and verify the validity of this universal relationship. More observations and theoretical models are required in order to completely understand the EoS of NS and to explore the possibilities of modified gravity theories for these objects.

\section*{Acknowledgements}
The author, K.N, would like to acknowledge BITS-Pilani for the international travel support and CFisUC, University of Coimbra, for their hospitality and local support during his visit in March 2023 for the purpose of conducting part of this research. T.M would like to thank FCT (Fundação para a Ciência e a Tecnologia, I.P, Portugal) under Projects No. UIDP/\-04564/\-2020, No. UIDB/\-04564/\-2020 and 2022.06460.PTDC for support.

\newpage

\bibliography{apssamp}

\end{document}